\documentclass[12pt]{iopart}
\usepackage{iopams}
\usepackage{graphicx}
\usepackage{esint}
\usepackage{subeqn}

\begin{document}

\title{Asymptotics of eigenvalues and eigenvectors of Toeplitz matrices}

\author{Hui Dai, Zachary Geary and Leo P. Kadanoff}

\address{The James Franck Institute, The University of Chicago,
929 E 57th Street, Chicago IL 60637, USA}
\ead{hdai@uchicago.edu, zgeary@mit.edu and l-kadanoff@uchicago.edu}

\begin{abstract}
A Toeplitz matrix is one in which the matrix elements are constant
along diagonals. The Fisher-Hartwig matrices are much-studied singular
matrices in the Toeplitz family. The matrices are defined for all
orders, $N$. They are parametrized by two constants, $\alpha$ and
$\beta$. Their spectrum of eigenvalues has a simple asymptotic form
in the limit as $N$ goes to infinity. Here we study the structure
of their eigenvalues and eigenvectors in this limiting case. We specialize
to the case with real $\alpha$ and $\beta$ and $0<\alpha<|\beta|<1$,
where the behavior is particularly simple.

The eigenvalues are labeled by an index $l$ which varies from $0$
to $N-1$. An asymptotic analysis using Wiener-Hopf methods indicates
that for large $N$, the $j$th component of the $l$th eigenvector
varies roughly in the fashion $\ln\psi_{j}^{l}\approx ip^{l}j+O(1/N)$.
The $l$th wave vector, $p^{l}$, varies as
\[
(I)\hspace{0.5in}p^{l}=2\pi l/N+i(2\alpha+1)\ln N/N+O(1/N)
\]
for negative values of $\beta$ and values of $l/(N-1)$ not too close
to zero or one. Correspondingly the $l$th eigenvalue is given by
\[
(II)\hspace{0.75in}\epsilon^{l}=a(\exp(-ip^{l}))+o(1/N)\hspace{0.5in}\
\]
 where $a$ is the Fourier transform (also called the {\em symbol})
of the Toeplitz matrix.

Note that $p^{l}$ has a small positive imaginary part. For values
of $j/N$ not too close to zero or one, this imaginary part acts to
produce an eigenfunction which decays exponentially as $j/N$ increases.
Thus, the eigenfunction appears similar to a bound state, attached
to a wall at $j=0$. Near $j=0$ this decay is modified by a set of
bumps, probably not universal in character. For $j/N$ above $0.6$
the eigenfunction begins to oscillate in magnitude and shows deviations
from the exponential behavior.

The case of $0<\alpha<\beta<1$ need not be studied separately. It
can be obtained from the previous one by a {}``conjugacy'' transformation
which takes $\psi_{j}$ into $\psi_{N-j-1}$. This {}``conjugacy''
produces interesting orthonormality relations for the eigenfunctions.

\end{abstract}

\vspace{2pc}
\maketitle

\tableofcontents{}

\section{Introduction}

\subsection{Definition}

A $N\times N$ Toeplitz matrix $T_{jk}$ is of the form
\begin{equation}
T=\left[\begin{array}{ccccc}
T(0) & T(-1) & T(-2) & T(-3) & \cdots\\
T(1) & T(0) & T(-1) & T(-2) & \cdots\\
T(2) & T(1) & T(0) & T(-1) & \cdots\\
T(3) & T(2) & T(1) & T(0) & \cdots\\
\vdots & \vdots & \vdots & \vdots & \ddots\end{array}\right]_{N\times N}\label{eq:ToeplitzMDef}
\end{equation}

\noindent A family of these matrices can be generated by doing a Fourier
transformation of a function $a(z)$ defined on the unit circle
\begin{equation}
T_{jk}=T(j-k)=\frac{1}{2\pi i}\ointctrclockwise_{S_{1}}dz\frac{a(z)}{z^{j-k+1}}\label{ToeplitzFT}
\end{equation}
 The generating function $a(z)$ is known as the symbol.

A Toeplitz operator is a Toeplitz matrix in which $N$, the number
of rows and columns is taken to be infinite. As $N$ goes to infinity,
the eigenvalues and eigenfunctions of the Toeplitz matrix might be
expected to converge to those of the corresponding operator. However,
the convergence is quite non-uniform and subtle as one can see from
the literature \cite{BS,TE}.

\subsection{History}

Toeplitz matrices have many applications in physics \cite{FF}. One
example is that in the two-dimensional Ising model, the spin-spin
correlation function of the square lattice can be written as a Toeplitz
determinant \cite{MPW,MW,LPK}. More generally they arise whenever
a line of {}``impurities'' exists in an otherwise uniform system.
This line is then represented by a matrix in which the interaction
between different impurities depends only on the distance between
them.

There has been a considerable study of the behavior of these matrices
in the limit as their order goes to infinity. Szeg\"o \cite{Sz} found
an asymptotic determination of Toeplitz determinants, which was then
extended to a wider class of models by Hartwig and Fisher \cite{FH1,FH2}.
The Szeg\"o class is defined by symbols which are non-singular on the
unit circle. These probably have no singularities in their eigenfunctions.
The more interesting class defined by Hartwig and Fisher has a symbol
of the form:
\begin{equation}
a_{\alpha,\beta}(z)=(2-z-1/z)^{\alpha}(-z)^{\beta}\label{eq:HF}
\end{equation}
 They have singularities involving irrational powers of $N$ in their
determinant. We shall find similar singularities in their eigenfunctions.

This work on singular Toeplitz matrices was in turn extended to give
the spectrum of eigenvalues \cite{BS,Widom1,Widom2}. Lee,
Dai, and Bettleheim \cite{SYL+HD+EB} did the first calculation of
the actual first order correction to the eigenvalue which they found to
be of order $(\ln N)/N$ when $l/N$ was neither very close to zero
or one. Their work was limited to the case $\alpha=0$. We shall extend
their eigenvalue calculation to use in our determination of the eigenfunctions.

\subsection{Outline of paper}

Our study is focused on the asymptotic behavior of the eigenvalues
and eigenvectors of the $N$ by $N$ Toeplitz matrices. The next chapter
is devoted to finding the qualitative properties of the large-$N$
eigenvectors, calculated numerically. We note that the eigenstates
can be classified by a momentum variable, $p^{l}$, which defines
the exponential behavior of the eigenvector. We also note that the
scaling properties of the eigenvectors at the ends of this interval
differ from the ones in the middle. We focus upon the latter.

Chapter 3 is devoted to the justification of a kind of quasi-particle
\cite{LFL,BP} theory for these eigenvalues and eigenvectors. Specifically
we develop arguments and numerical data for the two equations given
in the abstract. Both of these concern the {}``momentum'', $p^{l}$.
One of these defines an expression for the eigenvalue in terms of
the momentum and $N$. The other says that the momenta are roughly
equally spaced along the line $(0,2\pi)$. Both of these are derived
heuristically and checked by numerics.

Chapter 4 makes use of the Wiener-Hopf method to construct an analytic
theory of the eigenvectors. When $\alpha$ and $\beta$ are in the
right range, including for example, $0<\alpha<-\beta<1$, we can find
a closed form approximate expression for the eigenvectors appropriate
for small $j/N$. In Chapter 5, this integral expression is evaluated.
To carry out the evaluation, we must give, as input, the eigenvalue.
In this chapter, we show that, for large $N$, these operator eigenvectors
provide an excellent approximation to the corresponding matrix eigenvector
for values of $j/N<0.7$. Many of the scaling properties of the eigenvectors
and eigenvalues are derived from this analysis of the operators.

Chapter 6 briefly discusses these results and gives suggestions for
future work.

\section{The eigenvectors}

\subsection{Defining equations}

The Toeplitz matrices, of course, have $N$ eigenvalues corresponding
to $N$ right eigenvectors, $\psi_{j}^{l,N}$ and an equal number
of left eigenvectors $\tilde{\psi}_{j}^{l,N}$. (In discussing these
matrices, we shall most often drop the superscript, $N$.) The eigenvalues
and eigenvectors obey
\begin{subequations}
\begin{equation}
\sum_{j=0}^{N-1}T_{kj}\psi_{j}^{l,N}=\epsilon^{l,N}\psi_{k}^{l,N},{\rm ~for~all~}k\in[0,N-1]\label{eq:eigproblemr}
\end{equation}
 and
\begin{equation}
\sum_{j=0}^{N-1}\tilde{\psi}_{j}^{l,N}T_{jk}=\epsilon^{l,N}{\tilde{\psi}}_{k}^{l,N},{\rm ~for~all~}k\in[0,N-1].\label{eq:eigprobleml}
\end{equation}
\end{subequations}
Each of these eigenvectors are uniquely defined,
up to an overall multiplicative constant, whenever the eigenvalues
are non-degenerate.

\subsection{$N=\infty$}

For finite-$N$, we do not know closed form expressions for the eigenvalues
and eigenvectors of the Toeplitz matrix. However, the eigenvalues
and eigenvectors of the Toeplitz matrices are very easily found when
$T_{kj}$ is a doubly infinite matrix with indices covering $(-\infty,\infty)$.
Because of the translational invariance in the latter situation the
eigenvectors are of the form
\begin{equation}
\psi_{j}=\exp(ipj)\quad\mbox{and}\quad\tilde{\psi_{j}}=\exp(-ipj).\label{eq:infty-infty}
\end{equation}
 for all possible real values of $p$, while the corresponding eigenvalues
are determined by the symbol and are
\begin{equation}
\epsilon=a(\exp(-ip)).\label{eq:e-infty-infty}
\end{equation}

\subsection{Parity symmetry}

Note that for all finite Toeplitz matrices the right eigenvector can
be calculated from the left eigenvector by:
\begin{equation}
\tilde{\psi_{j}^{l}}=\psi_{N-1-j}^{l},{\rm ~for~all~}j\in[0,N-1]\label{eq:transpose}
\end{equation}
 Thus, for the finite Toeplitz matrix, $\tilde{\psi_{j}^{l}}$ serves
in an analogous role to that played by an adjoint eigenvector for
a Hermitian matrix. In particular, if $\tilde{\psi^{l}}$ and $\psi^{m}$
have different eigenvalues, they are orthogonal. We find that when
properly normalized, the vectors have orthonormality relations
\begin{subequations}
\begin{equation}
\sum_{j=0}^{N-1}~\tilde{\psi}_{j}^{l}\psi_{j}^{m}=\delta(l,m)\label{eq:ortho}
\end{equation}
 Before normalization, the vectors are arbitrary up to multiplication
by a complex constant. To get the normalization right one must multiply
by a constant, leaving only an ambiguity under multiplication by $\pm1$.
The completeness condition is
\begin{equation}
\sum_{l=0}^{N-1}~\tilde{\psi}_{j}^{l}\psi_{k}^{l}=\delta(j,k)\label{eq:complete}
\end{equation}
\end{subequations}
We have verified Eq.(\ref{eq:ortho}) and Eq.(\ref{eq:complete})
for the range of parameters considered in this paper.

The symbol's parity transform is produced by $z\rightarrow1/z$. For
the Fisher-Hartwig symbol of Eq.(\ref{eq:HF}), this change is reflected
by $\beta\rightarrow-\beta$. Thus in this case, $\beta$ and $-\beta$
are equivalent for finite $N$ since they are obtained from one another
by a symmetry operation. Thus they have the same eigenvalue spectrum.
For infinite $N$ they are not equivalent, so that in $j\in[0,\infty]$,
the matrices with parameters $\beta$ and $-\beta$ have different
eigenvalues.

\subsection{Numerical calculation of eigenvalues and eigenfunctions}

Our analysis begins by the calculation of the eigenvalues and eigenfunctions
of the Fisher-Hartwig finite-$N$ Toeplitz matrices. The calculations
are done in Mathematica. When ${\rm Re}(\alpha)>-\frac{1}{2}$, the
Toeplitz matrix elements are given by
\begin{eqnarray}
T_{jk}=T(j-k)=(-1)^{j-k}\frac{\Gamma(2\alpha+1)}{\Gamma(\alpha+1+\beta-j+k)\Gamma(\alpha+1-\beta+j-k)}\label{eq:Tjk}
\end{eqnarray}
These matrix elements have the asymptotic behavior
\begin{eqnarray*}
T(j-k) & \sim & -\frac{\Gamma(2\alpha+1)\sin\pi(\alpha\pm\beta)}{\pi(j-k+\beta)^{2\alpha+1}}{\rm ~~for~large~values~of~}|j-k|
\end{eqnarray*}
 Here the $\pm$ signs refer respectively to the cases in which $j>k$
and $j<k$.

In our numerical calculations, we use the exact form given by Eq.(\ref{eq:Tjk})
and then calculate eigenvalues and eigenfunctions using the routines
supplied in the commercial program Mathematica.

\begin{figure}
\begin{centering}
\includegraphics[height=8cm]{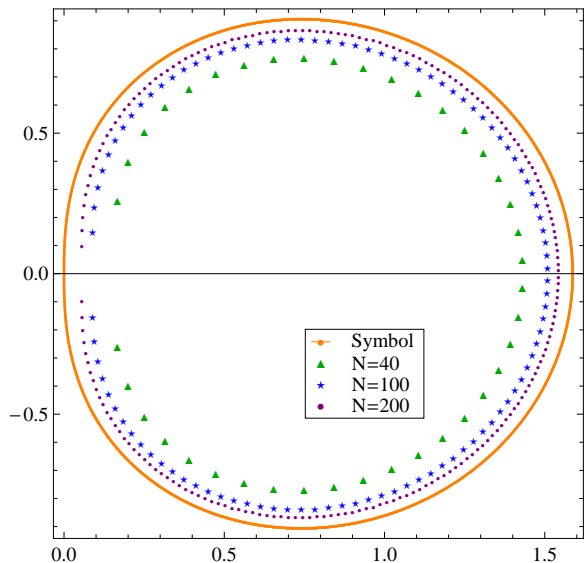}
\par\end{centering}

\caption{Eigenvalue Distribution. The image of the symbol together with the
eigenvalues in the complex plane for $\alpha=1/3$, $\beta=-1/2$;
the orange curve being the image of the symbol; the green triangles, blue stars and
purple dots being the eigenvalues for $N=40$, $100$ and $200$ respectively.}

\label{fig:ISymbol}
\end{figure}

To set the analysis of the eigenfunctions in a proper context we first
show in Figure \ref{fig:ISymbol} a depiction of the eigenvalues for
different $N$ and fixed $\alpha$ and $\beta$. The solid curve is
theoretical. It shows the image of the symbol, $a(z)$, as $z$ traverses
the unit circle. The dots are the eigenvalues calculated using Mathematica.
According to Widom's theory \cite{Widom1,Widom2}, for many kinds
of symbols, as $N$ goes to infinity, the spectrum of eigenvalues
should approach that image. Specifically, the eigenvalues are approximately
the ones given by the $N=\infty$ case in which fourier transformation
gives $\epsilon^{l,N}=a(\exp(-ip^{l}))$, where the momenta, $p^{l}$
are uniformly spaced in $(0,2\pi)$. In symbols,
\begin{equation}
\epsilon^{l,N}=a\left(e^{-2\pi il/(N-1)}\right)+O\left(\frac{\ln N}{N}\right)\label{eq:eigenmomentum}
\end{equation}
 This figure shows that, for the case plotted here, the eigenvalues
follow Widom's prescription. In fact Eq.(\ref{eq:eigenmomentum})
is true for all $\alpha$ and $\beta$ which obey $0<\alpha<|\beta|<1$,
which is the study range for this paper.

\subsection{Different regions for eigenvector}

\begin{figure}
\noindent \begin{centering}
\includegraphics[height=8cm]{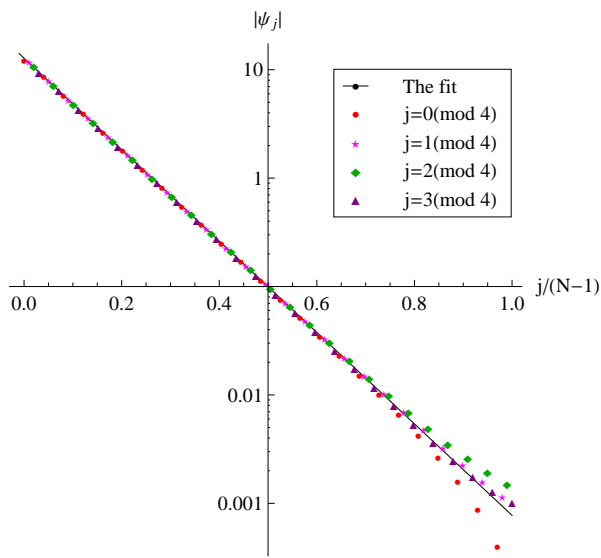}
\par\end{centering}

\caption{Magnitude of eigenvector. Here we plot $|\psi_{j}|$ against $j/(N-1)$
for $N=100$ and the eigenvalue with number $l=[(N-1)/4]$. If $l\sim pN/q$,
with $p$ and $q$ being integers the phase goes through roughly $p$
revolutions as $j$ increases by $q$. For this particular value of
$l$ there are four different branches to the eigenfunction and there
is a repetition after every four steps of $j$. The four branches
are indicated by different colors and shapes in this and the next figure. The
solid line is an exponential fit to the data. The parameter values
are $\alpha=1/3$ and $\beta=-1/2$. }

\label{fig:Mag}
\end{figure}

\begin{figure}
\noindent \begin{centering}
\includegraphics[height=8cm]{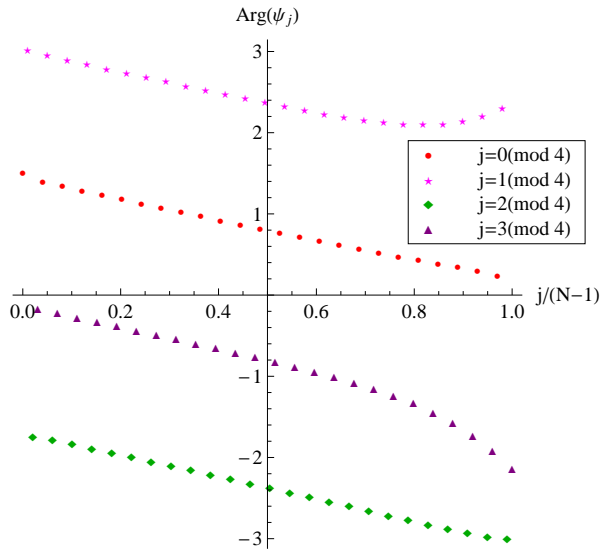}
\par\end{centering}

\caption{Phase of eigenvector. We plot the phase of the eigenvectors against
$j/(N-1)$. The overall behavior is that successive points increase
their phase by an amount close to $(2\pi)/4$. We again show the behavior
as four branches, each with points separated by four steps of $j$,
with the different branches being denoted by different colors and shapes. The
branches are separated by a height difference of about $\pi/2$. The
phase increases by an amount slightly smaller than $m\pi/2$ as $j$
advances by $m$ units. That is why the lines slope downward. }

\label{fig:Phase}
\end{figure}

The next two plots, Figure \ref{fig:Mag} and Figure \ref{fig:Phase}
show respectively the gross behavior of the magnitude and phase of
these eigenvectors, one with $N=100$ and $l=24$. Both plots show
that, for large $N$, the ratios $\psi_{j+1}/\psi_{j}$ are constant
throughout a broad central region of $j$, extending perhaps to from
small values of $j/(N-1)$ to $j/(N-1)\approx0.7$. This behavior
changes markedly within the first few values of $j$ and above $j/(N-1)\approx0.7$.
We call these two regions respectively the initial region and the
final region.

\subsection{Central region}

The first and most obvious characteristic of the central region is
that over most of the region the eigenvector varies exponentially
\begin{equation}
\psi_{j}^{l,N}\approx\exp(ip^{l,N}j)\label{eq:exponential}
\end{equation}
as indicated by the straight line behavior of exponent versus $j$
in Figure \ref{fig:Mag} and Figure \ref{fig:Phase}. This behavior
can most easily be understood as a consequence of the translational
invariance of the matrix $T_{j,k}=T(j-k)$. The central region is
far from the two ends $j=0$ and $j=N-1$. The eigenfunction in this
region behaves as it would if we were able to push the two ends infinitely
far away. We have picked $l=[(N-1)/4]$, so that this particular eigenfunction
shows four different branches which differ in phase by roughly $\pi/2$.

\subsection{Initial region}

\begin{figure}
\begin{centering}
\includegraphics[height=8cm]{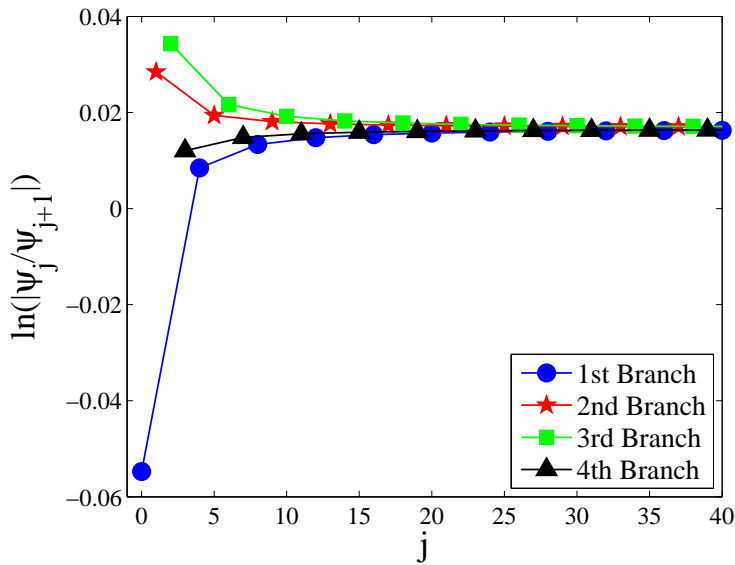}
\par\end{centering}

\caption{Magnitude of eigenvector in initial region. Here we plot four branches
of the eigenvector labeled by $k=0,1,2,3$. Each branch has $j=k({\rm mod}{\rm ~}4)$.
For each branch, $\ln|\psi_{j}/\psi_{j+1}|$, is plotted against $j$
for the eigenvalue with number $l=[(N-1)/4]$, $N=801$, and parameter-values
of $\alpha=1/3$, $\beta=-1/2$. The curves each show their different
individual behavior for small $j$ and then become almost identical
to one another for larger $j$. }

\label{fig:initialCompareBranches}
\end{figure}

\begin{figure}
\begin{centering}
\includegraphics[height=8cm]{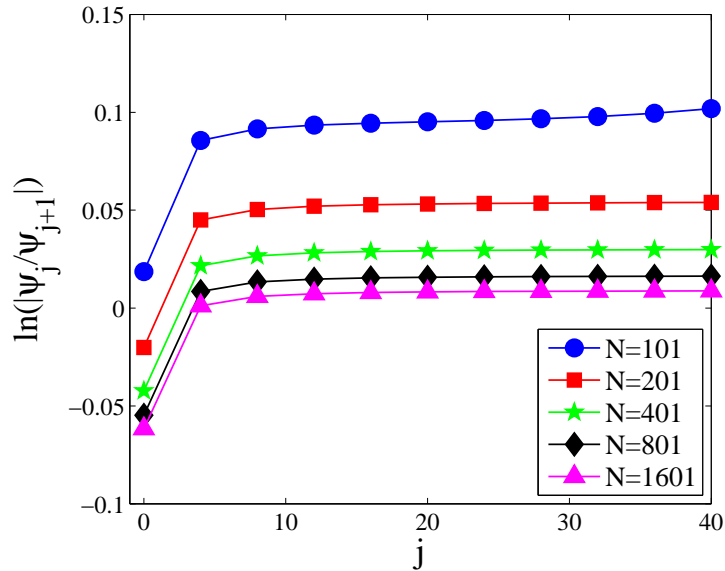}
\par\end{centering}

\caption{Magnitude of eigenvector in initial region for one branch. Here we
plot $\ln|\psi_{j}/\psi_{j+1}|$, against $j=4k$ for $N=101,201,401,801$
and $1601$ for the eigenvalue with number $l=[(N-1)/4]$ and parameter-values
of $\alpha=1/3$, $\beta=-1/2$. One can see that the shape of the curves
does not change appreciably as $N$ varies. So the initial region contains
eigenvectors with $N$-independent shapes. }

\label{fig:initialOneBranch}
\end{figure}

This branch-structure can also be discerned in the initial region
(See Figure \ref{fig:initialCompareBranches}.) In this plot, each
branch shows a few bumps for small values of $j$. The bumps decrease
in size as $j$ increases and the branches approach one another in
magnitude. (Their phases differ by about $(2\pi)/4$.) The behavior
then settles down to the exponential decay seen in the central region.
The behavior in the initial region does not appear to vary with $N$
for large values of $N$. Different values of $N$ show $\psi$ versus
$j$ plots which have, for small values of $j$, exactly the same
shape (See Figure \ref{fig:initialOneBranch}.)

\subsection{Final region}

\begin{figure}
\begin{centering}
\includegraphics[height=8cm]{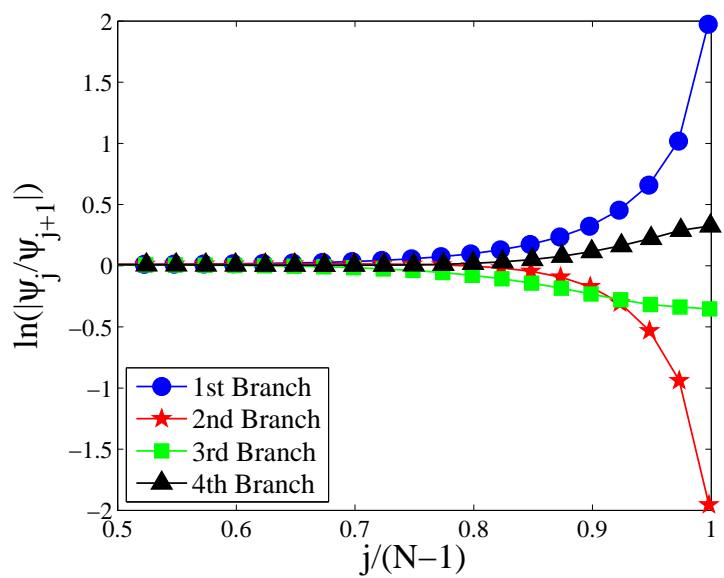}
\par\end{centering}

\caption{Magnitude of eigenvector in ending region. Here we plot $\ln|\psi_{j}/\psi_{j+1}|$,
against $j$ for $N=1601$ and the eigenvalue with number $l=[(N-1)/4]$
and parameter-values of $\alpha=1/3$, $\beta=-1/2$. The four-branch
structure is clearly indicated. }

\label{fig:final-branch}
\end{figure}

\begin{figure}
\begin{centering}
\includegraphics[height=8cm]{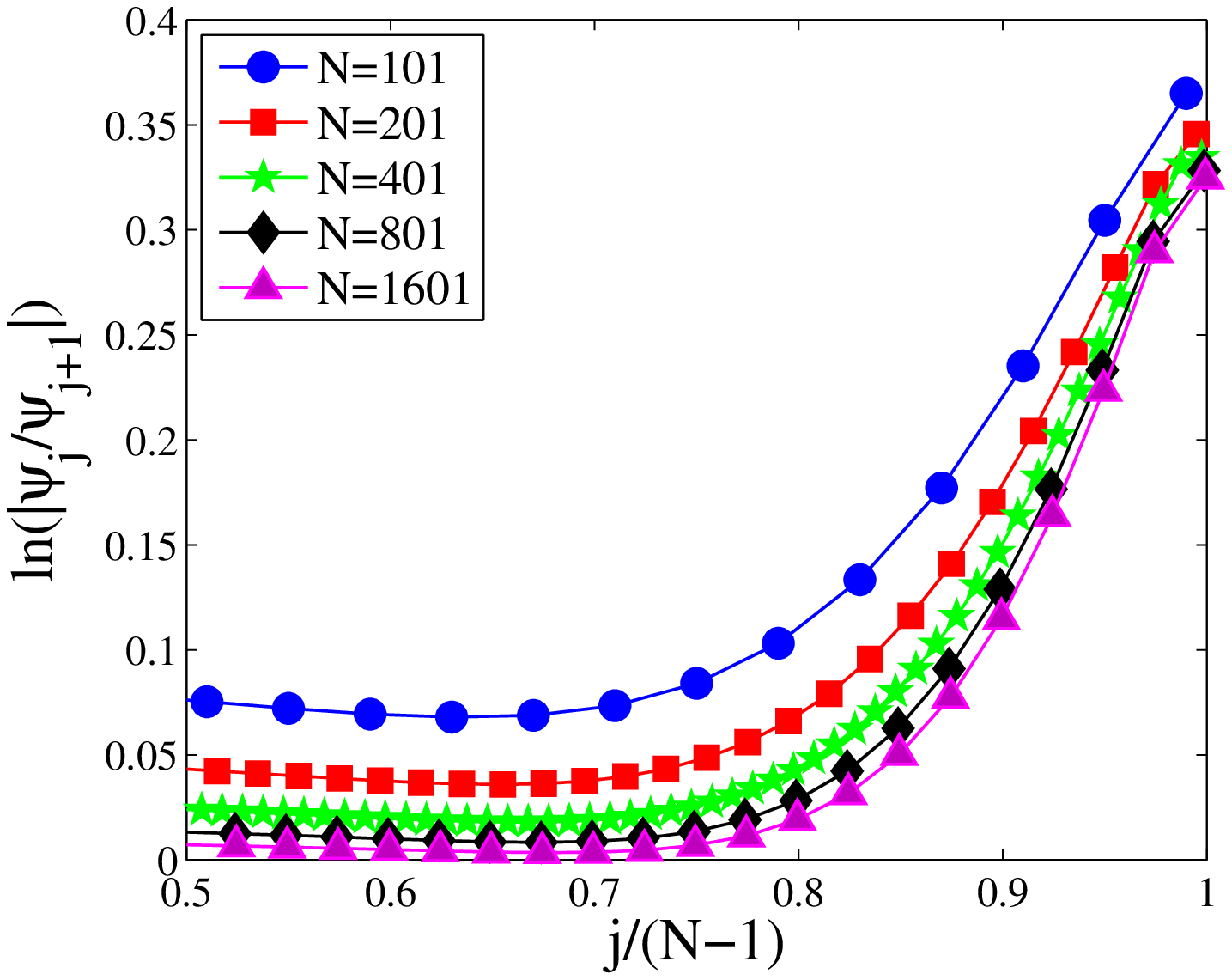}
\par\end{centering}

\caption{Magnitude of eigenfunction in ending region for one branch. Here we
plot $\ln|\psi_{j}/\psi_{j+1}|$, against $j=4k$ for $N=101,201,401,801,1601$
and the eigenvalue with number $l=[(N-1)/4]$ and parameter-values of
$\alpha=1/3$, $\beta=-1/2$. One can see that the shape of the curves
does not change much as $N$ varies. }

\label{fig:final-shape}
\end{figure}

Figure \ref{fig:final-branch} and Figure \ref{fig:final-shape} show
eigenfunction for relatively large values of $j/(N-1)$. We note from
the former figure that the four branches have different shapes in
this region. However, Figure \ref{fig:final-shape} shows that the
shapes of the different branches settle down to an $N$-independent
form as $N$ goes to infinity. However, the scaling in this region
appears to be somewhat complex. We leave a further analysis of this region
to a later publication.

\section{Quasi-particle results}

In the 1950s, Landau \cite{LFL,BP} worked out a theory of a low-temperature
Fermi liquid, e.g. He$^{3}$, which argued that the Fermi liquid was
just like a non-interacting Fermi gas except that it had a different
relation between energy and momentum. Here, we would like to make
the same kind of argument.

\subsection{{}``Energy''-momentum relation}

We follow Landau in saying that the relevant {}``momentum'', $p$,
appears in the phase of the eigenfunction as $\exp(ipj)$. In the
idealized case, $N=\infty$, corresponding to Landau's non-interacting
case, the phase and eigenvalue obey $\epsilon=a(\exp(-ip))$. We carry
this result over directly to finite values of $N$ as in Equation
II of the Abstract. We write that equation in terms of $a^{-1}(\cdot)$,
the function inverse to $a(\cdot)$, to define
\begin{subequations}\label{eq:def-p}
\begin{equation}
p_{\epsilon}^{l}=i\ln a^{-1}(\epsilon^{l})\label{eq:inverse-e}
\end{equation}
 The momentum variable defined from the wave function will be written
as $p_{\psi}$. One possible definition is
\begin{equation}
p_{\psi}^{l}=-i[\ln\psi_{J}^{l}-\ln\psi_{j}^{l}]/(J-j)\label{eq:p-psi}
\end{equation}
\end{subequations}
Here we must pick the branch of the logarithm
in Eq.(\ref{eq:def-p}) so as to make $p_{\psi}$ approximately equal
to $2\pi l/N$. We need to figure out whether, the two definitions
give results that are substantially equal in the large $N$ limit,
for example, whether it is true that
\begin{equation}
p_{\psi}^{l}=p_{\epsilon}^{l}+o(1/N)\label{eq:p-error}
\end{equation}

Eq.(\ref{eq:p-error}) and the equations in the abstract are all checked
using Tables 1 through 4. All four tables analyze the data for $\alpha=1/3$
and $\beta=-1/2$. The first two tables use data for the imaginary
part of $p$, while the latter two refer to the real part of $p$.
In each case, the $p_{\psi}$ in the table is obtained by using Eq.(\ref{eq:p-psi})
with $J=[0.5N]$ and $j=[0.2N]$. The first two lines in each table
describe the $p$-values thus obtained. They are seen to be increasingly
close for larger values of $N$. The next line gives the difference
between these values multiplied by powers of $N$. No one of these
lines increases substantially as $N$ increases. This result confirms
Eq.(\ref{eq:p-error}), which states that the two methods of determining
$p$ are consistent with one another in the limit of large $N$. In
fact these results tend to indicate that the difference between these
two estimates of $p$ is of order $1/N^{3}$ for the imaginary part
and $1/N^{2}$ for the real part. Thus we have verified Equation II
of the abstract.

\begin{figure}
\begin{centering}
\includegraphics[height=8cm]{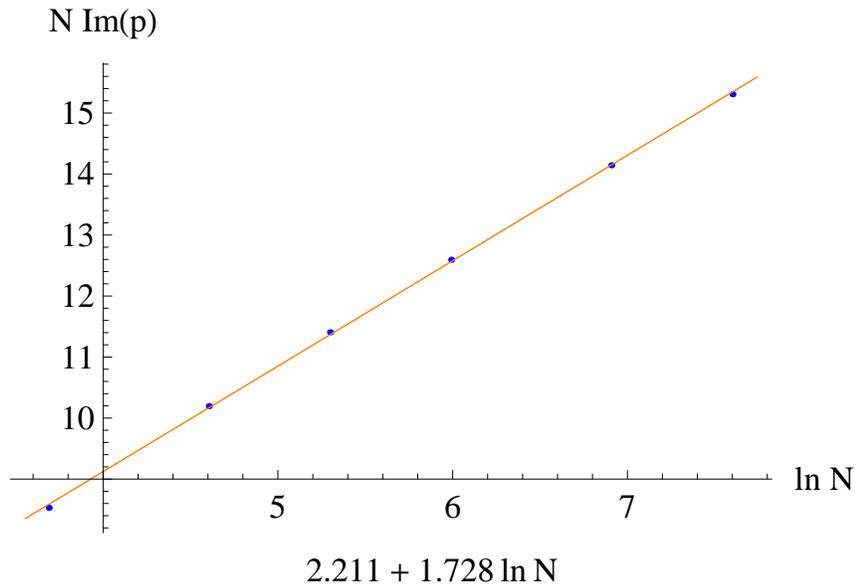}
\par\end{centering}

\caption{Imaginary part of $p$. This plot shows the data in the next to last
row of Table \ref{table 1} plotted along with a least squares fit.
The theoretical slope is $1.667$ and the fitted one is $1.728.$ }

\label{Fig:alpha=00003D1/3}
\end{figure}

Next we compare the $p$-values in the table with Equation I of the
abstract. The next to last row of the table gives the value of $N$
times the first row. According to Equation I, that row should be fitted
by $i(2\alpha+1)\ln N+C$, where $C$ is a complex constant. The final
row gives the errors for each $N$ after the corresponding theoretical
result is subtracted away. Each table thus involves one adjustable
constant. The small values of the remainder in the last row indicate
that Equation I of the abstract appears to be quite justified. The
fit for Table \ref{table 1} is shown in graphical form in Figure \ref{Fig:alpha=00003D1/3}.

\begin{table}
\begin{centering}
\begin{tabular}{|c|c|c|c|c|c|c|}
\hline
$N$ & $2000$ & $1000$ & $400$ & $200$ & $100$ & $40$\tabularnewline
\hline
\hline
$\Im(p_{\psi})$ & $0.00765381$ & $0.0141435$  & $0.0314841$  & $0.0570298$  & $0.101950$  & $0.213238$\tabularnewline
\hline
$\Im(p_{\epsilon})$ & $0.00765381$ & $0.0141435$  & $0.0314842$  & $0.0570311$  & $0.101961$  & $0.213463$\tabularnewline
\hline
difference$\cdot10^{6}$ & $-0.00111$ & $-0.00881$  & $-0.146$  & $-1.29$  & $-11.7$  & $-225$\tabularnewline
\hline
difference$\cdot N^{3}$ & $-8.85$ & $-8.81$  & $-9.36$  & $-10.3$  & $-11.7$  & $-14.4$\tabularnewline
\hline
\hline
$\Im(p_{\psi})\cdot N$ & $15.308$ & $14.146$  & $12.594$  & $11.406$  & $10.195$  & $8.530$\tabularnewline
\hline
$\Im(p_{\psi})\cdot N-{\rm Fit}$ & $0$ & $-0.00886$  & $-0.0316$  & $-0.0640$  & $-0.120$  & $-0.258$\tabularnewline
\hline
\end{tabular}
\par\end{centering}

\caption{Two estimates of the imaginary part of $p$ are shown to be substantially
equal. They are then successfully fit to the theory. The value of
$l$ for this table is $l=[(N-1)/2]$. The third line shows the difference
of the two estimates and the fourth line shows that difference multiplied
by $N^{3}$. The rough constancy of this line with increasing $N$
supports the correctness of Eq. (\ref{eq:p-error}). The next to last
line shows the result of multiplying the imaginary part of $p_{\psi}$
in the top row by $N$. According to the theory this quantity should
be $(2\alpha+1)\ln N+\Im(C)$. We get a good fit to these data by
using $\Im(C)=2.639$. The last row in the table is the result of subtracting
this fit from the result in the previous row. It represents the error
in our analysis. Since the relative error is so small, we argue that
Equation I of the abstract is corroborated. }

\label{table 1}
\end{table}

These results of this comparison are shown in Table \ref{table 1}
for $p$ roughly equals to $\pi/2$. Table \ref{table 2} shows the
same analysis applied to an {}``irrational'' value of $p$ chosen
by taking $l=[(\sqrt{5}-1)(N-1)/2]$. Tables \ref{table 3} and \ref{table 4}
give a similar comparison for the real part of $p$. In all cases
Eq.(\ref{eq:p-error}) appears to be justified.

\begin{table}
\noindent \begin{centering}
\begin{tabular}{|c|c|c|c|c|c|c|}
\hline
$N$ & $2000$ & $1000$ & $400$ & $200$ & $100$ & $40$\tabularnewline
\hline
\hline
$\Im(p_{\psi})$ & $0.00761559$ & $0.0140678$  & $0.0313017$  & $0.0566988$  & $0.101397$  & $0.212772$\tabularnewline
\hline
$\Im(p_{\epsilon})$ & $0.00761559$ & $0.0140678$  & $0.0313018$  & $0.0566981$  & $0.10141$  & $0.212644$\tabularnewline
\hline
difference$\cdot10^{6}$ & $0.000137$ & $0.00964$  & $-0.0699$  & $0.680$  & $-13.4$  & $128$\tabularnewline
\hline
difference$\cdot N^{3}$ & $1.10$ & $9.64$  & $-4.48$  & $5.44$  & $-13.4$  & $8.17$\tabularnewline
\hline
\hline
$\Im(p_{\psi})\cdot N$ & $15.231$ & $14.068$  & $12.521$  & $11.340$  & $10.140$  & $8.511$\tabularnewline
\hline
$\Im(p_{\psi})\cdot N-{\rm Fit}$ & $0$ & $-0.00811$  & $-0.0281$  & $-0.0538$  & $-0.0986$ & $-0.200$\tabularnewline
\hline
\end{tabular}
\par\end{centering}

\caption{The same as Table \ref{table 1} except that the values of $l$ are
given by $l=[(\sqrt{5}-1)(N-1)/2]$.}

\label{table 2}
\end{table}

\begin{table}
\begin{centering}
\begin{tabular}{|c|c|c|c|c|c|c|}
\hline
$N$ & $2000$ & $1000$ & $400$ & $200$ & $100$ & $40$\tabularnewline
\hline
\hline
$\Re(q_{\psi})\cdot10^{6}$ & $3.80031$ & $14.4177$  & $84.7912$  & $326.402$  & $1265.36$  & $7661.04$\tabularnewline
\hline
$\Re(q_{\epsilon})\cdot10^{6}$  & $1.72814$ & $6.91045$  & $43.1505$  & $172.352$  & $687.673$  & $4274.71$\tabularnewline
\hline
difference$\cdot10^{6}$  & $2.07$ & $7.51$  & $41.6$  & $154$  & $578$  & $3386$\tabularnewline
\hline
difference$\cdot N^{2}$  & $8.29$ & $7.51$  & $6.66$  & $6.16$  & $5.78$  & $5.42$\tabularnewline
\hline
\hline
$\Re(q_{\psi})\cdot N$  & $0.00760$ & $0.0144$  & $0.0339$  & $0.0653$  & $0.127$  & $0.306$\tabularnewline
\hline
$\Re(q_{\psi})\cdot N-{\rm Fit}$ & $0$ & $0.00682$  & $0.0263$  & $0.0577$  & $0.119$  & $0.299$\tabularnewline
\hline
\end{tabular}
\par\end{centering}

\caption{The same as Table \ref{table 1}, except that we examine here the
real part of $q=p-2\pi l/(N-1)$. The fit on the last line sets $N$
times the real part equal to a constant, $0.00760$. }

\label{table 3}
\end{table}

\begin{table}
\begin{centering}
\begin{tabular}{|c|c|c|c|c|c|c|}
\hline
$N$ & $2000$ & $1000$ & $400$ & $200$ & $100$ & $40$\tabularnewline
\hline
\hline
$\Re(q_{\psi})\cdot10^{6}$ & $-808.955$ & $-1610.85$  & $-4038.63$  & $-7861.35$  & $-14949$  & $-26243.8$\tabularnewline
\hline
$\Re(q_{\epsilon})\cdot10^{6}$ & $-810.805$ & $-1617.64$  & $-3997.34$  & $-7726.56$  & $-14391.3$  & $-29846.7$\tabularnewline
\hline
difference$\cdot10^{6}$ & $1.85$ & $6.80$  & $-41.3$  & $-135$  & $-558$  & $3603$\tabularnewline
\hline
difference$\cdot N^{2}$ & $7.40$ & $6.80$  & $-6.61$  & $-5.39$  & $-5.58$  & $5.76$\tabularnewline
\hline
\hline
$\Re(q_{\psi})\cdot N$ & $-1.618$ & $-1.611$  & $-1.615$  & $-1.572$  & $-1.495$  & $-1.050$\tabularnewline
\hline
$\Re(q_{\psi})\cdot N-{\rm Fit}$ & $0$ & $0.00707$  & $0.00246$  & $0.0456$  & $0.123$  & $0.568$\tabularnewline
\hline
\end{tabular}
\par\end{centering}

\caption{The same as Table \ref{table 2}, except that we examine here the
real part of $q=p-2\pi l/(N-1)$. The fit on the last line compares
$N$ times the real part equal to a constant, $-1.618$. As one can
see, this fit is quite accurate especially for the larger values of
$N$. }

\label{table 4}
\end{table}

\subsection{Spacing of eigenvalues}

The other part of quasi-particle theory that we wish to verify is
that for successive $l$-values the $p^{l}$ should be spaced by $2\pi/N$.
This will help ensure that the phase difference between $\psi_{0}^{l}$
and $\psi_{N-1}^{l}$ can remain independent of $l$. To check this
we notice that the fits for the real part of $p$ in Tables \ref{table 3}
and \ref{table 4} indicate that these two eigenvalues have their
spacing correctly described by the two equations in the abstract.
A brief examination shows that these spacings seem to continue to
hold for all values of $l$ except the ones which have $l/(N-1)\ll1$
or $1-l/(N-1)\ll1$. Near these end-points different spacings are
observed. In this paper, we do not pursue this point any further.

\section{Wiener-Hopf method}

\subsection{The singly infinite eigenvalue problem}

To get a good approximation for the Toeplitz matrix's eigenvector,
one might start by looking at the limiting case in which $N$ is infinity.
The case in which the integer index, $j$, is extended from $-\infty$
to $\infty$ is simply solved by Fourier transformation as in Eq.(\ref{eq:infty-infty})
and Eq.(\ref{eq:e-infty-infty}). But the singly infinite case, described
as the Toeplitz operator problem, in which $j$ extends from, say,
$0$ to $\infty$ is much harder. The approach from singly infinite-$N$
to the case of $N$ large but finite is tricky and non-uniform. (We
give an example of the non-uniformity below.) Nonetheless the singly
infinite matrix provides a good entry into the finite-$N$ system.

We want to solve the Toeplitz operator eigenvalue-problem
\begin{subequations}
\begin{equation}
\sum_{j=0}^{\infty}T_{kj}\psi_{j}=\epsilon\psi_{k},{\rm ~for~all~}k\geq0\label{eq:eigproblem}
\end{equation}
 and the transpose problem
\begin{equation}
\sum_{j=0}^{\infty}\tilde{\psi}_{j}T_{jk}=\epsilon\tilde{\psi}_{k},{\rm ~for~all~}k\geq0\label{eq:treigproblem}
\end{equation}
\end{subequations}
where $T$ is a Toeplitz operator, i.e. $T_{jk}=T(j-k)$.

There is a well-developed theory, called the Wiener-Hopf method, for
solving equations involving this kind of operator \cite{WH}. See for example \cite{Gakhov} and
Chapter IX of McCoy and Wu \cite{MW}. See also \cite{OT,ND} for
applications close to the problem studied here. Following McCoy and
Wu, we define a class of vectors with indices which run from $-\infty$
to $\infty$. The components of these special vectors are required
to vanish for negative values of their index. The vectors of this
type are indicated by the superscripts {}``$+$''. For example the
{}``$+$'' vector corresponding to an eigenfunction of the Toeplitz
operator is written as
\[
\psi_{j}^{+}=\left\{ \begin{array}{cc}
\psi_{j}, & j\geq0\\
0, & j<0\end{array}\right.
\]

\noindent then we can write the eigenvalue equation as
\[
\sum_{k=-\infty}^{\infty}K_{jk}\psi_{k}^{+}=0{\rm ~~for~all}~j\geq0~~~~{\rm where}~~K_{jk}=T_{jk}-\epsilon\delta_{jk}
\]

\noindent We would like to define an equation for the eigenfunction
which covers the entire range, $(-\infty,\infty)$, for the variable
$j$. To do that we define another kind of vector, one indicated
by a superscript {}``$-$''. One example of such a vector is \begin{equation}
\psi_{j}^{-}=\sum_{k=-\infty}^{\infty}K_{jk}\psi_{k}^{+},{\rm ~for~all~integer~}j\label{eq:-}\end{equation}
In general, the vectors in this category vanish for $j>0$. However
this particular vector also vanishes for index equal to zero. This
extra condition will play an important role in our solution.

The result of this subsection is an eigenvalue equation which reads
\begin{equation}
\sum_{k=-\infty}^{\infty}K_{jk}\psi_{k}^{+}=\psi_{j}^{-}\makebox{~~for all integer~~}j\label{eq:eigen-out}
\end{equation}
 This equation has the same content as Eq.(\ref{eq:eigproblem}).

\subsection{Fourier transforms}

\noindent Eq.(\ref{eq:eigen-out}) may be solved by Fourier transformation.
Define
\begin{eqnarray}
a(z) & = & \sum_{j=-\infty}^{\infty}T(j)z^{j}\label{eq:symbol}\\
\psi^{+}(z) & = & \sum_{j=-\infty}^{\infty}\psi_{j}^{+}z^{j}=\sum_{j=0}^{\infty}\psi_{j}^{+}z^{j}\label{eq:psiin}\\
\psi^{-}(z) & = & \sum_{j=-\infty}^{\infty}\psi_{j}^{-}z^{j}=\sum_{j=-\infty}^{-1}\psi_{j}^{-}z^{n}\label{eq:psiout}
\end{eqnarray}

Multiply Eq.(\ref{eq:eigen-out}) by $z^{j}$ and sum over all $j$,
to give
\[
\sum_{j=-\infty}^{\infty}\sum_{k=-\infty}^{\infty}(T_{jk}-\epsilon\delta_{jk})z^{j-k}\psi_{k}^{+}z^{k}=\sum_{j=-\infty}^{\infty}\psi_{j}^{-}z^{j}
\]

\noindent which then gives an eigenvalue equation expressed in Fourier
language:
\begin{equation}
K(z)\psi^{+}(z)=\left(a(z)-\epsilon\right)\psi^{+}(z)=\psi^{-}(z)\label{eq:alg}
\end{equation}
 At this point, the eigenvalue equation has been reduced to an algebraic
equation, one that may apparently be solved by algebraic manipulations.
However, Eq.(\ref{eq:alg}) has two unknowns, $\psi^{+}(z)$ and $\psi^{-}(z)$
so that additional knowledge is needed to obtain a solution
for $\psi^{+}(z)$. That knowledge is based upon the analytic properties
of the various functions.

\subsection{Factorization}

To find $\psi^{+}(z)$, we must factorize $K(z)=a(z)-\epsilon$ into
a product of functions of the form
\begin{equation}
K(z)=z^{\nu}\frac{K^{+}(z)}{K^{-}(z)}\label{eq:factor}
\end{equation}
 where $\nu$ is an integer determined by the form of $K(\cdot)$.
It is called the {\em winding number} since it represents the total
number of times the function wraps around zero when $z$ is rotated
once around zero in the positive sense. The winding number obeys
\begin{equation}
\nu=\frac{1}{2\pi i}\intop_{0}^{2\pi}d\ln K(e^{ip})=\frac{1}{2\pi i}\left[\ln K(e^{2\pi i})-\ln K(e^{0i})\right]\label{eq:winding}
\end{equation}
 This factorization is important since it will enable us to immediately
construct a non-trivial solution to the Toeplitz operator equation
in the one case, $\nu=-1$, in which this solution exists \cite{Gakhov}. (See McCoy
and Wu \cite{MW} page 208-215. The condition, $\nu=-1$, demands that
the $j=0$ component of $\psi_{j}^{-}$ must be zero, as is required
for our solution.)

A factorization like this can be derived simply when the function,
$K(z)$, has no singularities or zeros on the unit circle and has
none at zero or infinity. However, all these conditions fail in the
Fisher-Hartwig case. But, McCoy and Wu \cite{MW} argue that the
factorization also works with the definition, Eq.(\ref{eq:winding}),
of the winding number in the case in which $K(z)$ is only continuous,
not analytic, on the unit circle and has appropriate analyticity properties
in an analytic continuation away from that circle. Their analysis
then covers the Fisher-Hartwig case in the range interesting to us:
$0<\alpha<|\beta|<1$. If one applies this integral formula of Eq.(\ref{eq:winding})
to find the winding number for $K$ one sees that
\begin{subequations}
\begin{equation}
\nu=-1\mbox{~~ for~~}-1<\beta<0,\label{eq:nu=-1}
\end{equation}
so this is the case in which the Toeplitz operator gives a non-trivial
solution for the eigenvector. Conversely
\begin{equation}
\nu=1\mbox{~~ for~~}0<\beta<1
\end{equation}
\end{subequations}
so there are only trivial eigenvector solutions
in this situation.

McCoy and Wu also give us formulas for finding the functions $K^{\pm}(z)$.
One starts from the contour integration
\begin{equation}
G^{\pm}(z)=\frac{1}{2\pi i}\ointctrclockwise_{S_{1}}dz'\frac{\ln(z'^{-\nu}K(z'))}{z'-z} \label{eq:G}
\end{equation}
 Here the upper (and lower) signs respectively refer to the cases
in which $z$ is inside (and outside) the unit circle. Then
\begin{equation}
K^{\pm}(z)=\exp[G^{\pm}(z)]
\end{equation}

With these definitions, the eigenvalue equation has the form
\begin{equation}
\psi^{+}(z)K^{+}(z)=z^{-\nu}\psi^{-}(z)K^{-}(z)\label{eq:eigen-fact}
\end{equation}
 The left hand side contains only non-negative powers of $z$. To
say when we can have eigenvalues, we must look at the right hand side
and see whether it can match the left. The three factors on the right
hand side contains powers which are at a maximum $-\nu-1+0$, obtained
from reading the three factors from left to right.

\subsection{$\nu=-1$}

The sole possibility of obtaining a solution to Eq.(\ref{eq:eigen-fact})
comes from picking $\nu=-1$. Then the right hand side can match the
left by simply having each side be a constant independent of $z$.
Thus one finds
\begin{equation}
\psi^{+}(z)K^{+}(z)=z\psi^{-}(z)K^{-}(z)=C\label{eq:}
\end{equation}
 where $C$ is a non-zero constant which will set the phase and magnitude
of the wave function.

In this case, we have as our solution
\begin{equation}
\psi^{+}(z)=\frac{C}{K^{+}(z)}{\rm ~~for~~}|z|<1\label{eq:nu_-1_result}
\end{equation}
 We have now obtained a solution in terms of quadratures.

Note that in general we expect that the $\nu$-value will remain constant
as the eigenvalue varies over some open set. Thus one has a region
in the complex plane in which the Toeplitz operator has eigenvalues.
This result is in contrast to the outcome for the Toeplitz matrix
which usually has $N$ isolated eigenvalues.

\subsubsection*{Example: A simple pole}

We choose a very simple example to illustrate what we have said. Take
$a$ to be of the Fisher-Hartwig type with $\alpha=0$ and $\beta=-1$:
$a(z)=z^{-1}$ (See Eq.(\ref{eq:HF}), where for convenience we have
dropped the overall minus sign). The matrix, $T_{jk}$, is
\begin{equation}
T^{\infty}=\left[\begin{array}{ccccc}
0 & 1 & 0 & 0 & \cdots\\
0 & 0 & 1 & 0 & \cdots\\
0 & 0 & 0 & 1 & \cdots\\
0 & 0 & 0 & 0 & \cdots\\
\vdots & \vdots & \vdots & \vdots & \ddots\end{array}\right]\label{eq:T-infinity}
\end{equation}
 The equation $\phi=T\psi$ is equivalent to \[
\phi_{j}=\psi_{j+1}~~\makebox{for}~~j\ge0.\]
 Thus the operator $T=z^{-1}$ when it is acting to the right transfers
information from large $j$ to smaller $j$. It is reasonable to assume
that all operators with $\nu=-1$ similarly transfer information towards
smaller $j$, and thereby they can effectively produce a boundary
condition of a vector being zero as $j$ goes to infinity.

Take the eigenvalue to be $\epsilon$, with magnitude less than unity,
so that it sits within the unit circle. Then \[
K(z)=z^{-1}-\epsilon=K^{+}/(zK^{-})\]
 The factorization then gives \[
K^{+}(z)=1-\epsilon z\mbox{~~and~~}K^{-}(z)=1.\]
 As required $K^{+}$ has its zero outside the unit circle. One therefore
finds that the eigenfunction has a fourier transform \[
\psi^{+}(z)=\frac{C}{1-\epsilon z}=C\left(1+\epsilon z+(\epsilon z)^{2}+\cdots\right),\]
 which in turn implies an exponentially decaying eigenfunction
 \begin{equation}
\psi_{j}=C(\epsilon)^{j}\makebox{~~for~~}j=0,1,2,\cdots
\end{equation}
 Thus the entire unit circle is filled with eigenvalues while $\psi^{+}(z)$
has a single pole, and the eigenfunctions $\psi_{j}$ are geometrical
series.

This situation is of the Fisher-Hartwig type with $\alpha=0$ and
$\beta=-1$.

However, the finite Fisher-Hartwig matrix with this symbol has a very
different behavior. This matrix, $T^{(N)}$ has $N$ unit entries
just above the diagonal, with all the rest of the entries being zero.
(Here the notation $T^{(N)}$ describes an $N$ order Toeplitz matrix
and not the $N$th power of anything.) Thus it looks just like the
infinite matrix of Eq.(\ref{eq:T-infinity}) except that it is truncated
after $N$ rows and $N$ columns, i.e.
\begin{equation}
T^{(N)}=\left[\begin{array}{ccccc}
0 & 1 & 0 & 0 & \cdots\\
0 & 0 & 1 & 0 & \cdots\\
0 & 0 & 0 & 1 & \cdots\\
0 & 0 & 0 & 0 & \cdots\\
\vdots & \vdots & \vdots & \vdots & \ddots\end{array}\right]_{N\times N}\label{xxx}
\end{equation}
 The eigenvalue equations read
\begin{subequations}
\begin{equation}
\psi_{N-1}=0
\end{equation}
 and
\begin{equation}
\psi_{N-1-j}=(1/\epsilon)\psi_{N-j}\makebox{~~for~~}j=1,2...N-1
\end{equation}
\end{subequations}
These equations are solved successively. They
do not make sense for $\epsilon=0$. For all other values of $\epsilon$,
the eigenvector has all of its elements being zero. As a result, all
eigenfunctions are trivial.

Thus, the eigenfunction behavior of this particular order-$N$ Fisher-Hartwig
matrix matches that of the corresponding operator only in the special
case in which we set $C=0$ in the operator eigenvector.

\section{Calculation of eigenfunctions}

The work of previous authors on eigenvalues has given us useful but
limited information. We have an asymptotic expansion which gives the
line on which the eigenvalues fall and their probability density on
that line. Their exact placement on that line has never been explicitly
calculated. Nor have the eigenfunctions associated with these eigenvalues.
Here, we set out to fill in some of these omissions, using both asymptotic
calculations and numerical work. We shall especially make use of the
Wiener-Hopf solutions outlined in the previous chapter to discuss
the properties of the behavior arising from the Fisher-Hartwig generating
function.

\begin{figure}
\noindent \begin{centering}
\includegraphics[height=8cm]{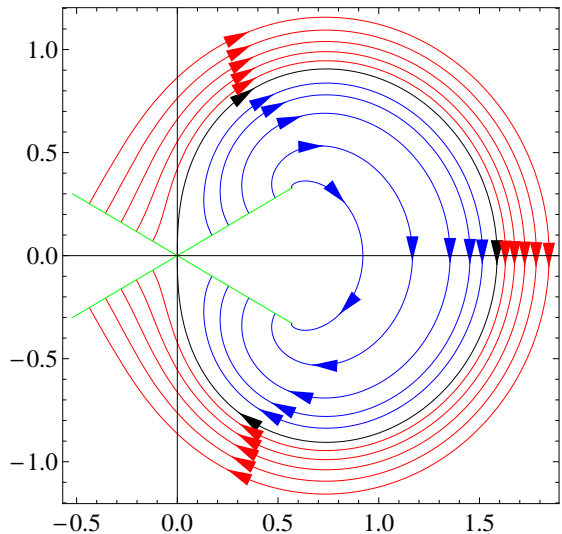}
\par\end{centering}

\caption{Maps of the symbol on the $w$-plane. The black curve is the image
of the symbol $w=a_{\alpha,\beta}(z)$ for $\alpha=1/3$ and $\beta=-1/2$
with $z$ going around the unit circle in the counterclockwise direction.
The symbol has a singularity at $z=1$ (mapped to $w=0$) as we can
see from the other curves in this figure. The curves counting from
the outmost are the images of this symbol for circles $|z|=0.75$,
$|z|=0.8$, $|z|=0.85$, $|z|=0.9$, $|z|=0.95$, $|z|=1$, $|z|=1.1$,
$|z|=1.2$, $|z|=1.4$, $|z|=2$ and $|z|=4$ respectively (red curves
for $|z|<1$, blue curves for $|z|>1$ and the black curve for the
unit circle). The green lines consist of the endpoints of the curves.
\label{fig:analytic-symbol}}

\end{figure}

\begin{figure}
\begin{centering}
\includegraphics[height=8cm]{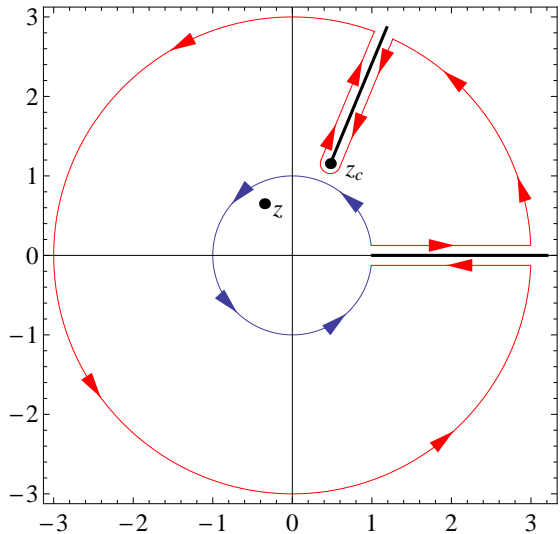}
\par\end{centering}

\caption{The singularities in the contour integral of in the $z'$-plane for
the integration in Eq.(\ref{eq:G+ for FH}). The original contour,
on the unit circle, is shown in blue. This path can then be deformed
into the one shown in red, encircling singularities outside the unit
circle. The black lines are encircled branch lines, including one
starting at $z_{c}$, where $a(z')=\epsilon$, and another along the
positive real axis. The path on the big red circle can be taken to
infinity where it does not contribute. The dot is the singularity
at $z'=z_{c}$. There is an additional pole singularity at $z'=z$, which
does not contribute because $z$ is inside the unit circle. }

\label{fig:deformed}
\end{figure}

For $0<\alpha<-\beta<1$, all the eigenvalues lie within the closed
curve formed as the image of the unit circle mapped by the symbol.
Since the image winds around in the clockwise direction, the
winding number of $(a_{\alpha,\beta}-\epsilon)$ is $-1$ for any
eigenvalue, $\epsilon$. Therefore, we can use the quadrature of the
previous chapter,

\begin{equation}
\psi^{+}(z)=Ce^{-G^{+}(z)}\label{eq:psiin for FH}
\end{equation}

\noindent where
\begin{equation}
G^{+}(z)=\frac{1}{2\pi i}\ointctrclockwise_{S_{1}}dz'\frac{\ln[z'(a_{\alpha,\beta}(z')-\epsilon)]}{z'-z}\label{eq:G+ for FH}
\end{equation}

\noindent This result is then a solution by quadratures of the Toeplitz
operator problem involving the Fisher-Hartwig symbol. We may assume
that this solution is accurate for small positive values of $x=j/(N-1)$
and becomes less accurate as $x$ approaches one from below.

\subsection{Evaluation of the integrals}

Eq.(\ref{eq:psiin for FH}) and Eq.(\ref{eq:G+ for FH}) together
give the result for the Fisher-Hartwig symbol for the case $0<\alpha<-\beta<1$.
Figure \ref{fig:ISymbol} shows the image of the unit circle formed
by the symbol together with the eigenvalues of the corresponding Toeplitz
matrices for the case $\alpha=1/3$, $\beta=-1/2$. We see that the
eigenvalues are all distributed inside the image of the symbol. They
approach that image as the size of the Toeplitz matrix goes to infinity.
Further discussion of the asymptotic behavior of the eigenvalues has
been given in \cite{SYL+HD+EB}.

To perform the contour integration in Eq.(\ref{eq:G+ for FH}), we
need to rewrite our symbol in terms of an analytic function, which
is
\begin{equation}
a_{\alpha,\beta}(z)=(z-1)^{2\alpha}z^{\beta-\alpha}e^{-i(\beta+\alpha)\pi}\label{eq:analytic-symbol}
\end{equation}

We shall choose this function, which has branch points at zero, one,
and infinity, to have a branch cut along the positive real axis (See
Figure \ref{fig:deformed}). The logarithm in Eq.(\ref{eq:G+ for FH})
has an additional singularity at the point, outside the unit circle,
where $a(z')$ equals the eigenvalue. We call this point $z_{c}$.
It obeys
\begin{equation}
a(z_{c})=\epsilon\label{eq:zc}
\end{equation}
We can then have another branch line extending from $z_{c}$ to $\infty$.
As $N$ goes to infinity, $\epsilon$ approaches the image of the
unit circle and $z_{c}$ approaches the unit circle. These points
will play important roles in what follows. Since $a(z')$ goes exponentially
to zero at infinity the contour in Eq.(\ref{eq:G+ for FH}) can be
deformed as shown in Figure \ref{fig:deformed}. The big circular
path can then be taken to $\infty$. The result is that our integral
becomes, aside from an additive constant,

\begin{subequations} \label{eq:quadrature}
\begin{equation}
G^{+}(z)=L(z)+S(z)
\end{equation}
 Here $L(z)$ stands for a logarithmic term, which is
\begin{equation}
L(z)=\ln(z-z_{c})
\end{equation}
 while $S(z)$ stands for a term which has a weak singularity at $z=1$,
\begin{equation}
S(z)=\frac{1}{2\pi
i}\intop_{1}^{\infty}\frac{dt}{t-z}\ln\left[\frac{(t-1)^{2\alpha}t^{\beta-\alpha}e^{-i(\beta+\alpha)\pi}-\epsilon}{(t-1)^{2\alpha}t^{\beta-\alpha}e^{i(\beta+\alpha)\pi}-\epsilon}\right]\label{eq:Sint}
\end{equation}
\end{subequations}

Here, $z_{c}$ is the position of the zero in $K(z)=a(z)-\epsilon$
in the $z$-plane. According to Widom's ideas, this zero approaches
the unit circle as $N$ goes to infinity, but remains outside of that
circle. This approach makes for a near-singularity in $\psi^{+}(z)$
on the circle. For $z$ inside the circle, this near-singularity looks
like a singularity just outside the point $z=\exp(-2\pi il/(N-1))$.
In contrast the real singularity in $\psi^{+}(z)$ is produced by
the branch line which passes through $z=1$. The singularity and the
near-singularity are well separated except for situations in which
$l$ is relatively close to $l=0$ or $l=N-1$. We have not explored
these exceptional limiting cases.

\subsection{Accuracy of Wiener-Hopf solution }

\begin{figure}
\noindent \begin{centering}
\includegraphics[height=8cm]{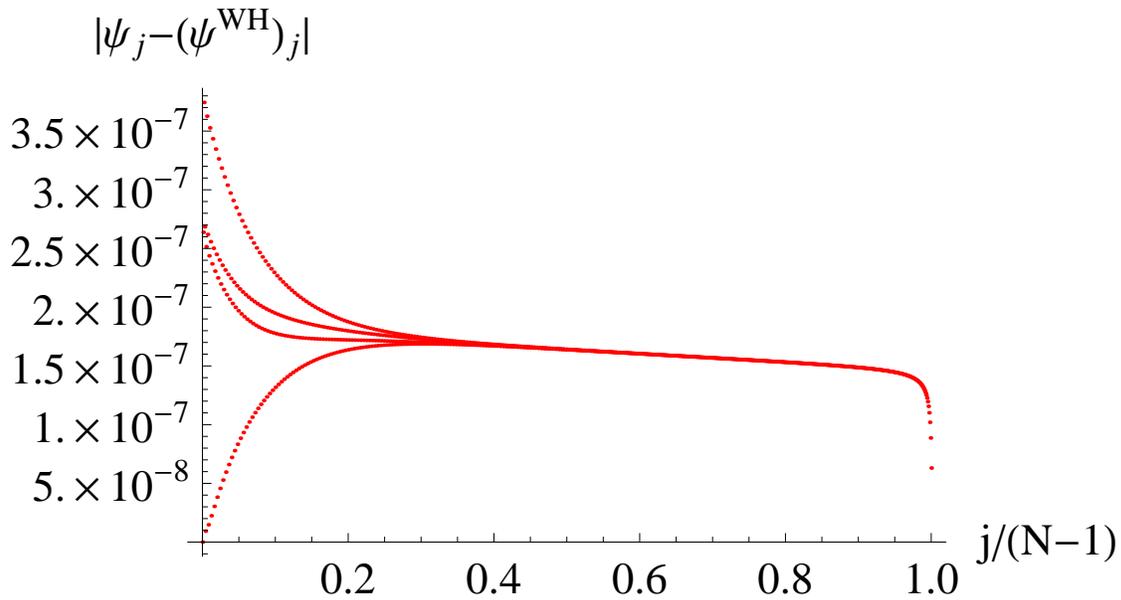}
\par\end{centering}

\caption{Absolute comparison of Wiener-Hopf solution with exact eigenfunction.
The magnitude of the difference between the Wiener-Hopf solution and
the exact eigenfunction is plotted against $j/(N-1)$ for $N=1000$, $l=[(N-1)/4]$,
$\alpha=1/3$ and $\beta=-1/2$. }

\label{fig:compare-absolut}
\end{figure}

\begin{figure}
\noindent \begin{centering}
\includegraphics[height=8cm]{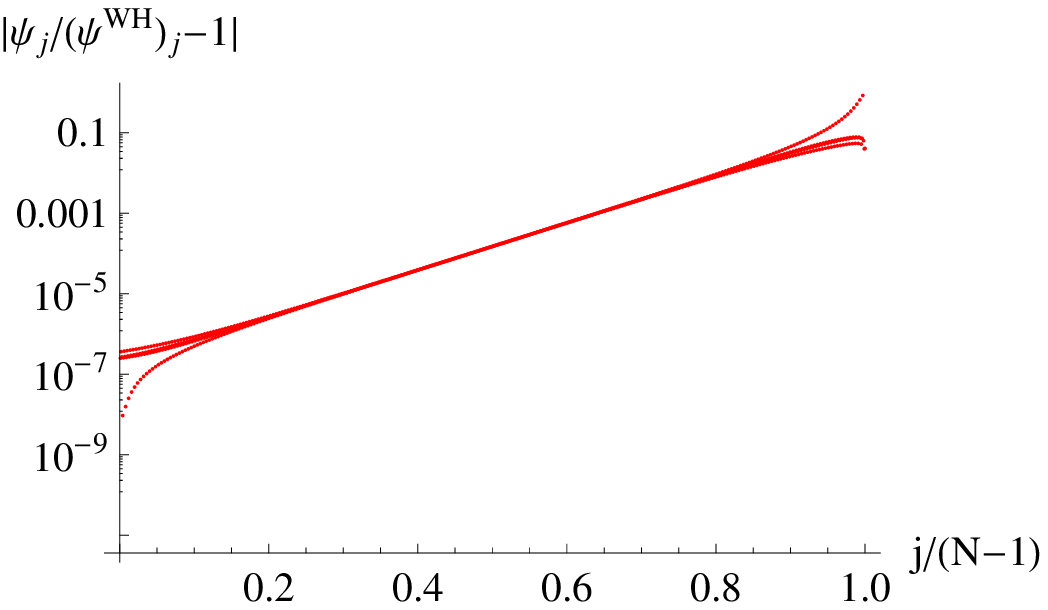}
\par\end{centering}

\caption{Relative comparison of Wiener-Hopf solution with exact eigenfunction.
The relative magnitude, $\left|\psi/\psi^{WH}-1\right|$ of the difference
between the Wiener-Hopf solution and the exact eigenfunction is plotted
against $j/(N-1)$ for $N=1000$, $l=[(N-1)/4]$, $\alpha=1/3$ and $\beta=-1/2$. }

\label{fig:compare-relativ}
\end{figure}

The solution we have generated is exact in the limiting case in which
$j$ is fixed and $N$ goes to infinity. Thus, we might expect that
the solution is accurate for large $N$ and fixed $j/(N-1)$. To see how
accurate it is, we plot in Figure \ref{fig:compare-absolut} the absolute
value of the difference between the Wiener-Hopf quadrature and the
exact solution for $N=1000$. The absolute value of the error never
gets larger than a few times $10^{-7}$. It decreases slowly as $j/(N-1)$
increases. The relative error, plotted in Figure \ref{fig:compare-relativ},
grows with $j/(N-1)$, starting from order $10^{-7}$ and increasing to
order one at $j/(N-1)\approx1$. Therefore, we may say that the Wiener-Hopf
solution reproduces the main features of the exact eigenfunction except
very close to the maximum values of $j/(N-1)$.

\subsection{Examination of solution}

Even a superficial examination of Eq.(\ref{eq:quadrature}) shows
that its behavior fits the known properties of the eigenfunction.
One term in the wave function is \begin{equation}
\psi_{L}(z)=\exp\left(-L(z)\right)=-\frac{1}{z_c(1-z/z_c)}=-\frac{1}{z_c}\sum_{j=0}^{\infty}\left(z/z_{c}\right)^{j}\end{equation}
 This part of $\psi(z)$ thus produces a $\psi_{j}$ which is \begin{equation}
(\psi_{L})_{j}=-(1/z_{c})^{j+1}\label{eq:exponential1}\end{equation}
 This structure is exactly the decaying exponential so evident in
the central region of $j$.

The other parts of $\psi(z)$ come from $\psi_{S}(z)=\exp\left(-S(z)\right)$.
This exponential is independent of $N$ except for the very weak $N$-dependence
produced by a very small additive term of the form $\gamma(\ln N)/N$
in the eigenvalue, $\epsilon$. This effect of the $N$-dependence
in this term is quite negligible.

\begin{figure}
\begin{centering}
\includegraphics[height=8cm]{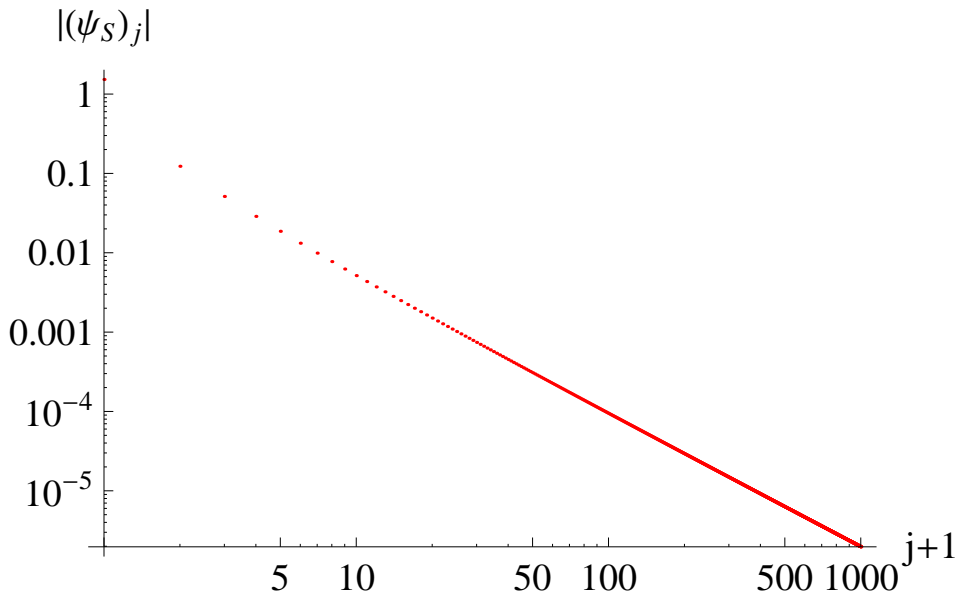}
\par\end{centering}

\caption{Expansion of $\psi_S$. The magnitude of the terms in the expansion of
$\psi_S(z)$ in powers of $z$. These terms reflect corrections to the
eigenfunction beyond the leading exponential behavior. The notable
features are a long tail proportional to $(j+1)^{-2\alpha-1}$ and a
few decreasing order-one bumps for small $j$. }

\label{fig:S-expand}
\end{figure}

Figure \ref{fig:S-expand} plots the magnitude of the terms in the
power series expansion of $\psi_{S}(z)$ in $z$. Using the same notation
as used for $\psi_{L}(z)$, these terms are called $(\psi_{S})_{j}$.
The first few terms are of order unity and produce the small-$j$
bumps evident in Figure \ref{fig:initialCompareBranches}. The bumps
slowly die away for higher $j$.

For higher $j$ the behavior of $(\psi_{S})_{j}$ is dominated by
a power law behavior, which shows up as a straight line in Figure
\ref{fig:S-expand}. Even though these terms fall off, this contribution
remains significant. This behavior is shown in Eq.(\ref{eq:power}).
It results from a singularity as $z$ goes to unity evidenced in Eq.(\ref{eq:Sint}),
which shows a behavior for small $z-1$ of the form
\[
\ln\psi_{S}(z)\sim-\frac{\sin\pi(\alpha+\beta)}{\pi~\epsilon}\int_{0}^{L}~ds~\frac{s^{2\alpha}}{1+s-z}
\]
$L$ is a large number which serves as a cutoff at large values of
$s$. This expression can be expanded in a power series in $z$, with
coefficients of high order terms in the form
\[
\left[\ln\psi_{S}(z)\right]_{j}\sim-\frac{\sin\pi(\alpha+\beta)}{\pi~\epsilon}\int_{0}^{L}~ds~\frac{s^{2\alpha}}{(1+s)^{j+1}}
\]
For large $j$ the integral contributes for small values of $s$,
of order $1/j$ so that we can replace $(1+s)^{j+1}$ by $e^{s(j+1)}$.
We thus get the leading singularity in $\ln\psi_{S}(z)$ to be given
by
\begin{equation}
[\ln\psi_{S}(z)]_{j}\sim-\frac{\Gamma(2\alpha+1)\sin\pi(\alpha+\beta)}{\pi~\epsilon~(j+1)^{2\alpha+1}}\label{eq:algebraic}\label{eq:power}
\end{equation}
Since this singularity in $\ln\psi_{S}(z)$ indicates a zero added
onto a leading term which is a constant, the very same form of singularity
appears in $\psi_{S}(z)$, which therefore has a high order expansion
\begin{equation}
[\psi_{S}(z)]_{j}\sim-\frac{\Gamma(2\alpha+1)\sin\pi(\alpha+\beta)}{\pi~\epsilon~(j+1)^{2\alpha+1}}\label{eq:algebraic}
\end{equation}
 for large values of $j$.

\subsection{Mixed terms}

\subsubsection{Sums}

So far we have argued that the solution for $\psi_{j}$ should contain
terms of the exponential form shown in Eq.(\ref{eq:exponential1})
as well as the algebraic form in Eq.(\ref{eq:algebraic}). In addition
there is a set of bumps which appear only for small $j$. For larger
$j$, then, we might expect an eigenfunction which looks like
\begin{equation}
\psi_{j}\sim A(z_{c})^{-j-1}+B(j+1)^{-(2\alpha+1)}\label{eq:sum}
\end{equation}
 where $A$ and $B$ are both constants. Since both terms are generated
near $j=0$, we expect both $A$ and $B$ to be of order unity. Note
that both terms fall off for higher $j$. If, as we expect, $\Re\left(\ln1/z_{c}\right)$
is small, then the exponential decay will be slow for small $j$,
while the decay of the algebraic term will be rapid. However, for
sufficiently large $j$ the two terms return to being of the same
order. We assume this equality is achieved when $j=O(N)$ so that
we might have a boundary condition in which the two terms effectively
interfere to produce a very small result near and beyond the border
at $j=N-1$:
\[
(z_{c})^{-N}\sim N^{-(2\alpha+1)}
\]
 Since $z_{c}=\exp(-ip)$, we can calculate the imaginary part of
$p$ as
\begin{equation}
\Im\left(p\right)=\Re\left(\ln z_{c}\right)=(2\alpha+1)\frac{\ln N}{N}+O(1/N)
\end{equation}
 The $O(1/N)$ term arises from the ratio of $A/B$ in Eq.(\ref{eq:sum}).
This change in $p$ enables one to calculate the shift in the eigenvalue
in the form
\[
\frac{\delta\epsilon}{\epsilon}=\frac{\partial\ln\epsilon}{\partial p}\Im\left(p\right)
\]
 so that the change in the eigenvalue is given by
\begin{equation}
\frac{\delta\epsilon}{\epsilon}=\left(\beta+i\alpha\cot\frac{p}{2}\right)(2\alpha+1)\frac{\ln N}{N}\label{eq:eigenvalue}
\end{equation}
 This is, to our knowledge, the first calculation of the eigenvalue
shift for $\alpha$ not equal to zero.

\subsubsection{Products}

The actual form of the Wiener-Hopf eigenfunction is not a sum of the
form shown in Eq.(\ref{eq:sum}) but instead a product in which
\begin{equation}
\psi(z)\sim(z-z_{c})^{-1}\psi_S(z)\label{eq:prod-z}
\end{equation}
which can be Fourier transformed to give
\begin{equation}
\psi_{j}\sim\frac{\Gamma(2\alpha+1)\sin\pi(\alpha+\beta)}{\pi~\epsilon~z_{c}}\sum_{k=0}^{j}z_{c}^{-j+k}\frac{1}{(k+1)^{2\alpha+1}}\label{eq:prod-j}
\end{equation}
We introduce an integral representation for the last factor to give
\[
\psi_{j}\sim\frac{\sin\pi(\alpha+\beta)}{\pi~\epsilon~z_{c}}\int_{0}^{\infty}~d\mu~\mu^{2\alpha}\sum_{k=0}^{j}e^{-(k+1)\mu}z_{c}^{-j+k}
\]
The geometric series can be summed to find
\[
\psi_{j}\sim \frac{\sin\pi(\alpha+\beta)}{\pi~\epsilon}\left[\int_{0}^{\infty}~d\mu~e^{-\mu}\mu^{2\alpha}\frac{z_{c}^{-j-1}-e^{-\mu (j+1)}}{1-z_{c}e^{-\mu}}\right]
\]
This elegant result enables us to write $\psi_{j}$ as a sum of two
terms:
\begin{subequations}
\begin{equation}
\psi_{j}\sim Az_{c}^{-j-1}+B(j+1)^{-(2\alpha+1)}\label{eq:prod-result}
\end{equation}
where the term multiplied by $A$ dominates in the central region
and both terms are of the same order in the ending region. Here the
value of the coefficients are
\begin{equation}
A\sim\frac{\sin\pi(\alpha+\beta)}{\pi~\epsilon}\int_{0}^{\infty}~d\mu~e^{-\mu}\mu^{2\alpha}\frac{1}{1-z_{c}e^{-\mu}}
\end{equation}
 and
\begin{equation}
B\sim-\frac{\sin\pi(\alpha+\beta)}{\pi~\epsilon}\int_{0}^{\infty}~d\mu~e^{-\mu}\mu^{2\alpha}\frac{1}{1-z_{c}e^{-\mu/(j+1)}}
\end{equation}
\end{subequations}
To construct this form for $B$ we have made the
change of variables $\mu\rightarrow\mu/(j+1).$ Note that this expression
gives an eigenfunction of precisely the same form as we have analyzed
in Eq.(\ref{eq:sum}) in the previous subsection. Therefore the previous
analysis will hold, if we assume that the coefficients $A$ and $B$
are constants. The only problem is that the integrals for $A$ and
$B$ have singularities if $z_{c}$ is close to unity. For large $N$
this value of $z_{c}$ will only occur if $l/(N-1)$ or $\left(1-l/(N-1)\right)$ is very small. If we are not in those special regions of $l$,
then $A$ and $B$ can be considered to be constants of order unity
so that our previous estimate of the eigenvalue given by Eq.(\ref{eq:eigenvalue})
will hold.

\section{More to investigate}

This paper is only a partial investigation of the problem at hand.
It focuses on obtaining a good solution for smaller values of $j/(N-1)$
and does not look in detail at values of $j/(N-1)$ very close to one. In particular,
the form of large-$j$ scaling is not considered. One might expect
that some sort of Wiener-Hopf solution might be obtained on the large-$j$
side and that values of $p^{l}$ and $\epsilon^{l}$ might be calculable
through order $1/N$.

No calculations are extended to the case $\alpha=0$, nor to the case
in which $|\beta|<\alpha$.

We have shed no light on the source of the Fisher-Hartwig $\alpha^{2}-\beta^{2}$
in the determinant of $T$ nor have we explained what happens when
$l/(N-1)$ is close to zero or one.

Nonetheless this paper does contain a large number of new results:
we see excellent eigenfunctions for smaller $j$, and corrections
in $p^{l}$ and $\epsilon^{l}$ to order $(\ln N)/N$. More importantly
we have shed some light on how the logarithms arise to permit the
interference between an exponential and an algebraic term in the eigenfunction.
In all of this, we have made a novel use of {}``quasi-particle''
theory a la Landau.

\section*{Acknowledgments}

We would like to thank Michael Fisher, Peter Constantin, and Harold
Widom for helpful discussions. This research was supported by the
University of Chicago MRSEC under grant number NSF-DMR 0820054. Zachary
Geary particularly wishes to thank the Chicago MRSEC-REU program for
supporting his stay at Chicago.

\newpage{}


\begin{thebibliography}{16}
\bibitem{BS} B\"ottcher A and Silbermann B, \emph{Introduction to Large Truncated Toeplitz Matrices}, 1999
\emph{Springer}

\bibitem{TE} Trefethen L and Embree M, \emph{Spectra and Pseudospectra}, 2005
\emph{Princeton University Press}

\bibitem{FF} Forrester P and Frankel N E, \emph{Applications and Generalizations
of Fisher-Hartwig Asymptotics}, 2004 \emph{J. Math. Phys.} \textbf{45}
2003-2028

\bibitem{MPW} Montroll E W, Potts R B and Ward J C, \emph{Correlations and
spontaneous magnetization of the two-dimensional Ising model}, 1963
\emph{J. Math. Phys.} \textbf{4} 308-322

\bibitem{MW} McCoy B M and Wu T T, \emph{The Two-Dimensional Ising Model}, 1973
\emph{Harvard University Press}

\bibitem{LPK} Kadanoff L P, \emph{Spin-Spin Correlation in the Two-Dimensional
Ising Model}, 1966 \emph{Il Nuovo Cimento B} \textbf{44} 276-305

\bibitem{Sz} Szeg\"o G, \emph{Ein grenzwertsatz \"uber die Toeplitzschen Determinanten
einer reellen positiven Funktion}, 1915 \emph{Funktion. Math. Ann.} \textbf{76}
490-503

\bibitem{FH1} Fisher M E and Hartwig R E, \emph{Toeplitz determinants,
some applications, theorems and conjectures}, 1968 \emph{Adv. Chem. Phys.}
\textbf{15} 333-353

\bibitem{FH2} Fisher M E and Hartwig R E, \emph{Asymptotic behavior of
Toeplitz matrices and determinants}, 1969 \emph{Arch. Rat. Mech. Anal.} \textbf{32} 190-225

\bibitem{Widom1} Widom H, \emph{Toeplitz determinants with singular generating
functions}, 1973 \emph{Amer. J. Math.} \textbf{95} 333-383

\bibitem{Widom2} Widom H, \emph{Eigenvalue distribution of nonselfadjoint
Toeplitz matrices and the asymptotics of Toeplitz determinants in
the case of nonvanishing index}, 1990 \emph{Operator Theory: Adv. and Appl.}
\textbf{48} 387-421

\bibitem{SYL+HD+EB} Lee S Y, Dai H and Bettelheim E, \emph{Asymptotic
eigenvalue distribution of large Toeplitz matrices}, 2007 \emph{arXiv: 0708.3124v1}

\bibitem{LFL} See Wikipedia article on Fermi Liquid

\bibitem{BP} Baym G and Pethick C, \emph{Landau Fermi-Liquid Theory: Concepts
and Applications}, 2008 \emph{Wiley-VCH}

\bibitem{WH} Wiener N and Hopf E, \emph{\"Uber eine Klasse singul\"aer Integralgleichungen}, 1931
\emph{Sitz. Berlin. Akad. Wiss.} 696-706

\bibitem{Gakhov} Gokhov F D, \emph{Boundary Value Problems}, 1966 \emph{Pergamon Press, Addison-Wesley Pub. Co.}

\bibitem{OT} Ohtaka K and Tanabe Y, \emph{Theory of the soft-x-ray edge problem
in simple metals: historical survey and recent developments}, 1990 \emph{Rev. Mod. Phys.}
\textbf{62} 929-991

\bibitem{ND} Nozieres P and De Dominicis C T, \emph{Singularities in the X-Ray Absorption
and Emission of Metals. III. One-Body Theory Exact Solution}, 1969 \emph{Phys. Rev.} \textbf{178}
1097-1107

\bibitem{BM} Basor E and Morrison K E, \emph{The Fisher-Hartwig Conjecture
and Toeplitz Eigenvalues}, 1994 \emph{Linear Algebra Appl.} \textbf{202} 129-142
\end{thebibliography}
\end{document}